\documentstyle[aps,prl,multicol,graphicx]{revtex}
\begin{document}
\draft

\title{Valley Splitting in Si-Inversion Layers at Low Magnetic Fields}
\author{V.\ M.\ Pudalov$^a$, A.\ Punnoose$^b$,
G.\ Brunthaler$^c$, A.\ Prinz$^c$, G.\ Bauer$^c$}
\address{$^a$ P.\ N.\ Lebedev Physics
Institute, Moscow, Russia.}
\address{$^b$ Weizmann Institute of Sciences, Rehovot, Israel}
\address{$^c$ Institut f\"{u}r Halbleiterphysik,
Johannes Kepler Universt\"{a}t, Linz, Austria}
\maketitle

\begin{abstract}
We report novel manifestation of the valley splitting
for the two valley electron system in (100) Si-inversion layers
at low carrier density. We found that valley splitting causes
almost 100\% modulation of the Shubnikov de Haas
oscillations {\em in very low magnetic fields},
almost on the bound of the quantum
interference peak of the negative magnetoresistance.
From  the interference pattern of oscillations we determined
the valley splitting in the $B=0$ limit which appears to
vary  within a factor of 1.3
over the density range $(3-7)\times 10^{11}$\,cm$^{-2}$.
Within the same range of densities, level broadenings in
both electron valleys differ only
by $\leq 3\%$. The latter result shows that the inter-valley scattering is not
responsible for the strong (six fold) `metallic-like'
changes of the resistivity with temperature.

\end{abstract}

\pacs{PACS: 71.30.+h, 73.40.Hm, 73.40.Qv}

\begin{multicols}{2}
The apparent `metallic-like' temperature dependence of the
resistivity is  in the focus of current research interest. A number of
microscopic mechanisms have been proposed
for explaining the two major
features of the resistivity:
(i) strong  metallic-like changes  with temperature
\cite{MOSFET_MIT,other_MIT,hanein_9805113,hanein_9808251,simmons_9910368,ensslin},
and (ii)
strong changes with in-plane magnetic field
\cite{instability,simonian97,okamoto,yoon_9907128,disorder}.
One of the promising models intensively
discussed in this connection is
based on scattering between different sub-bands \cite{murzin,sivan,tutuc}.
If the carriers mobility strongly
depends on  the Fermi energy $E_F$, the minority subband(s)
may have a mobility essentially lower than that for the majority
subband(s). The intersubband scattering may thus cause
strong changes in the resistivity with  temperature and magnetic field.
In addition to  the spin-related
subbands,  $n-$(100)-Si system  has two minima in the
conduction band. The minima originate from six equivalent
valleys  located close to the $X$-points
in the Brillouin zone.
Four of them shift up in energy  by about (20-40)\,meV
due to the confinement potential.
At low densities,
only the two lowest valleys are filled;
the sharp Si/SiO$_2$ interface causes their
additional splitting by $\Delta_v \gtrsim 1$\,K\ \cite{ando}.
Therefore, apriory, the remarkable strength of the
metallic-like conduction in Si-samples might
be related to the valley
multiplicity.

The `metallic' temperature changes in the resistivity and
the `metal-insulator' transition (MIT) in 2D systems
\cite{MOSFET_MIT,other_MIT}  are most pronounced in high-mobility samples
where they take  place in the regime of low carrier density and strong
electron-electron interaction (the ratio of the Coulomb interaction energy
to Fermi energy, $r_s$, is of order of 3-10). The interaction is expected
to enhance both, spin- and valley-splitting \cite{ando}; a possibility of
a spontaneous polarization, caused by interactions, was discussed earlier
\cite{finkelstein} and was recently recalled again in connection with the
problem of the  MIT in 2D \cite{vitkalov_0009454}.

In {\em strong quantizing magnetic fields} $B_{\perp}$ ($\omega_c\tau \gg1$),
valley splitting is known to be enhanced by
exchange interaction between energy levels \cite{ando} and was directly
measured from  Shubnikov-de Haas (ShdH), quantum Hall (QHE) effect,
and chemical potential oscillations.
In particular, in Ref.~\cite{JETP85}, valley splitting
was found to be almost density independent and to increase
linearly with magnetic field,
$\Delta_v(B) = \Delta_v^0 + \alpha B$, where $\alpha=0.6$\,K/T and the sample dependent
$\Delta_v^0 \approx 2$\,K. The latter renormalization is
intrinsic only to the strong field regime of well separated energy levels
and is
irrelevant to the problem of metallic conduction
at $B=0$; valley splitting in  Si-structures at $B\rightarrow 0$
remained so far unexplored.

In the current work we have found that valley splitting manifests
itself in high mobility Si-samples not only at high fields,
but also {\em  at  low magnetic
fields, $B < 0.8$\,T}, giving rise to  beatings in the ShdH
oscillations. Such beatings were discussed theoretically
in Ref.~\cite{ando}\,b but have not been observed experimentally.
Due to very high mobility of the samples
we traced the oscillations down to the fundamental limit of $\approx 0.2$\,T, given by
the onset of the quantum interference
(i.e., weak negative magnetoresistance) \cite{weakloc}.
From this novel beating pattern we
determined the zero-field valley splitting as a function of the
carrier density. We found that
$\Delta_v$ varies only weakly in the range of low densities,
$n=(3 - 7)\times 10^{11}$\,cm$^{-2}$.
Almost 100\%   amplitude  of modulation in the beating pattern
evidences
that the
{\em scattering times
 in the two valleys are about equal}. This
observation demonstrates that the inter-subband scattering
mechanism is not responsible for the strong `metallic-like' temperature dependence of
the conduction in Si-inversion layers.

We performed measurements on three high mobility  $n-$Si-MOS
samples selected from three different
wafers (whose plane coincide with (100)-crystal plane
to within $1^{\rm o}$): Si11 (peak mobility
$\mu^{\rm peak}=3.9$\,m$^2$/Vs\ \ at $T=0.3$\,K), Si12  ($\mu^{\rm peak} =3.4$)
and Si15  ($\mu^{\rm peak} =4.0$).
All samples exhibited strong
(six-fold) metallic-like fall in the resistivity with temperature,
and an apparent MIT at a 'critical density' $n_c \approx 0.9\times
10^{11}$\,cm$^{-2}$.

Figure~1 shows a picture
of the oscillations typical for  high magnetic fields and for a
relatively high carrier density,
$n=10.24\times 10^{11}$\,cm$^{-2}$. As the magnetic field increases,
ShdH oscillations
evolve into the quantum Hall effect. In quantizing fields
the energy spectrum
is
\begin{equation}
\varepsilon = \hbar \omega_c (N+ \frac{1}{2})
              \pm \frac{1}{2} g^*\mu_B B
              \pm \frac{1}{2}\Delta_v(B),
\end{equation}
where  $g^*$ is the effective $g-$factor, $\mu_B$ is the Bohr magneton,
and $N \geq 0$ is the Landau level index.

\vspace{0.35in}
\begin{figure}
\begin{center}
\includegraphics[angle=0,width=3.2in]{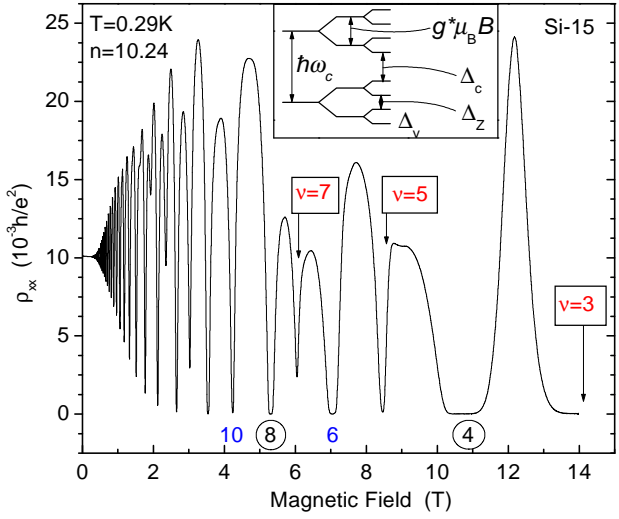}
\begin{minipage}{3.2in}
\vspace{0.1in} \caption{Typical manifestation of valley splitting
in diagonal resistivity at {\em high  fields and high carrier
density.} Arrows point at the valley gaps $\nu=5,7$, encircled
numbers depict filling factors for the cyclotron gaps, and
$\nu=6, 10$ are for the Zeeman
gaps. Inset illustrates the hierarchy of the energy levels in high fields.}
\label{fig1}
\end{minipage}
\end{center}
\end{figure}
\vspace{-0.2in}

The inset to Fig.~1\ illustrates
the energy
spectrum ~Eq.~(1)  in the high field regime. The
strongest minima in Fig.~1\,a (and the largest splittings) at high
fields correspond to the cyclotron gaps $\Delta_c=\hbar \omega_c -
g^*\mu_B B -\Delta_v$; they occur at integer filling factors $\nu =4, 8,
12...$, where $\nu= nh/eB_{\perp}$. The series of minima at $\nu
=2, 6, 10,$ etc. corresponds to Zeeman-gaps, $\Delta_z = g^*\mu_B
B - \Delta_v$. Starting from $B=6$\,T,  valley gaps
become resolved and produce minima in $\rho_{xx}$ at $\nu=7, 5,
3, 1$. For magnetic fields less than 6\,Tesla,  valley splitting
is not seen in Fig.~1\  at any
density, to within $\sim 1$\% resolution.

Spin splitting ($\nu=6,10, 14,$ etc) can be seen in Fig. 1
down to about 1.5\,T.
For
lower
fields, the oscillations decay exponentially in
amplitude, become
harmonic and obtain the anticipated
periodicity $4Be/hn$ corresponding to four degenerate levels (2
spins and 2 valleys).

Unexpectedly,
in the  low field range, $B=0.2-0.3$\,T, almost on the border with
the quantum interference peak in $\rho_{xx}(B)$  \cite{weakloc}, we found
beatings in the oscillatory picture, demonstrated in Figs.~2.
The beatings are observed
in all three samples but
only at low densities
$n<9 \times 10^{11}$\,cm$^{-2}$. For
$n<3\times 10^{11}$\,cm$^{-2}$,
the decreasing number of oscillations in the interference pattern
prevented us from making a quantitative analysis.

\vspace{0.1in}
\begin{figure}
\begin{center}
\includegraphics[angle=0,width=3.1in]{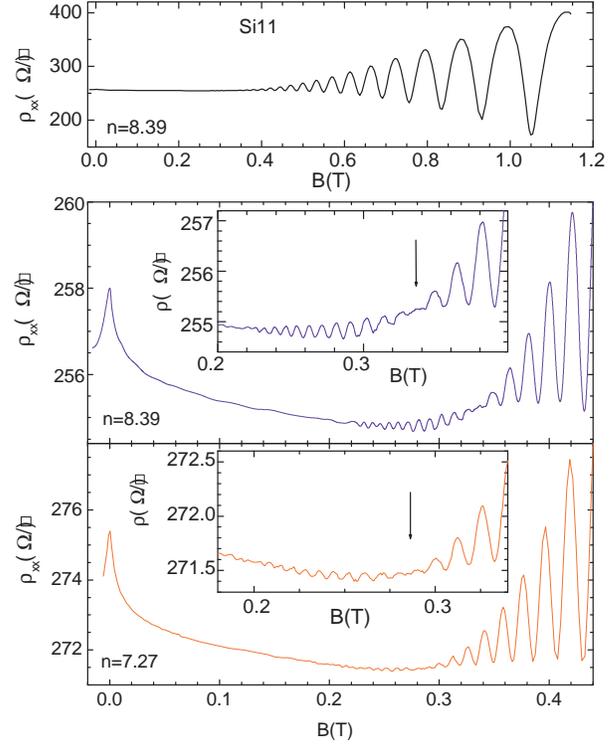}
\begin{minipage}{3.2in}
\vspace{0.1in}
\caption{Shubnikov-de Haas oscillations {\em at low
magnetic fields and low densities}: a)over an extended range of
fields, for $n=8.39\times
10^{11}$cm$^{-2}$ ; b) and c) magnified low field region of the interference pattern,
at two carrier densities, $n=8.39$ and $7.27\times
10^{11}$\,cm$^{-2}$. Arrows
mark  the beating nodes.}
\label{fig2}
\end{minipage}
\end{center}
\end{figure}
\vspace{-0.15in}

The interference picture in Figs.~2  corresponds obviously to the
presence of two oscillatory  terms in $\rho_{xx}(B^{-1})$ with
slightly different periods and almost the same amplitude. We note,
firstly, that Zeeman splitting in purely perpendicular field  in
no case may produce such interference. Secondly,
since both, cyclotron and Zeeman energy vanish as $B_{\perp}$ decreases,
valley splitting should become the dominant parameter in low
magnetic fields.
Indeed,
based on the measurements  of $\Delta_v$ \cite{JETP85} and
of $g^*$-factor \cite{gm} we estimate
\begin{eqnarray}
\Delta_c = \hbar\omega_c -\Delta_Z - \Delta_v
           \approx 1.8 B_{\perp}-2, \nonumber\\
\Delta_Z = g^*\mu_B B_{\perp} -\Delta_v \approx  2B_{\perp} - 2 \nonumber\\
\Delta_v \approx 2+ 0.6 B_{\perp},
\end{eqnarray}
where all energies are in K, $B_{\perp}$ is in T,
$\hbar\omega_c=7\times(0.19/m^*) \approx 5$\,K/T
\cite{ando,gm}, and $g^*\mu_B B \approx 2.6$\,K/T \cite{gm} for a
typical density  $n=5\times 10^{11}$\,cm$^{-2}$.

We
presume therefore that the interference
is caused by
valley splitting.
With this presumption, we modeled the oscillatory component of the
magnetoresistance $\delta\rho_{xx}(B_\perp)$ with
Lifshitz-Kosevich (LK) formulae  \cite{lifshitz_55,isihara_86}
for spin-degenerate carriers:
\begin{eqnarray}
\frac{\delta\rho_{xx}(B_{\perp})}{\rho(0)}=
2\sum_s A_{s}^{+} \cos \left( 2\pi s\frac{h n_{v+}}{2eB_{\perp}}- \pi s \right)+ \nonumber\\
A_{s}^{-} \cos \left( 2\pi s\frac{h n_{v-}}{2eB_{\perp}}- \pi s \right).
\end{eqnarray}
Here $n_{v\pm} = (n/2)\left[ 1 \pm
\Delta_v/(2\varepsilon_F)\right]$ are the partial populations in
the $\pm$ valleys. The envelope function is
\begin{equation}
A_s^{\pm} =  \exp\left(\frac{-\pi s}{\omega_c
\tau_q^{\pm}}\right) \frac{2\pi^2 skT/\hbar
\omega_c^*}{\sinh(2\pi^2 skT/\hbar \omega_c^*)}
F_s(g^*m^*),
\end{equation}
with $\tau_q^{\pm}= \hbar/(2\pi T_D^{\pm})$ being the quantum life
time, and $T_D^{\pm}$, the Dingle temperature in each $\pm$
valley. The Zeeman factor, $F=\cos(\pi s g^*\mu_B
B/\hbar\omega_c)= \cos(\pi s g^*m^*/2m_0)$ \cite{lifshitz_55},
and $m^*(n)$, $m_0$, are the effective and free electron masses
correspondingly.

\vspace{0.1in}
\begin{figure}
\begin{center}
\includegraphics[angle=0,width=3.1in]{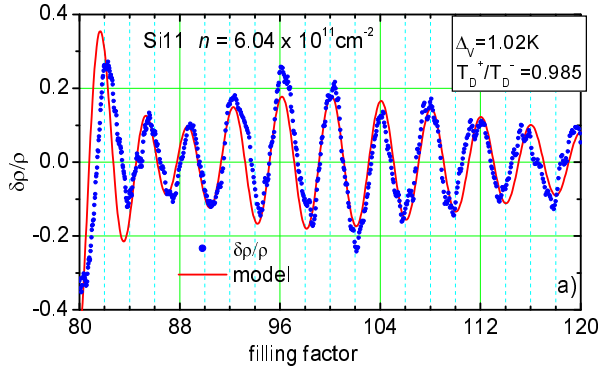}
\begin{minipage}{3.2in}
\vspace{0.1in}
\caption{Example of fitting of the measured interference pattern with
Eq.~(2) for the sample Si11. $n=6.04\times 10^{11}$\,cm$^{-2}$;
$T= 0.3$\,K.}
\label{fig3b}
\end{minipage}
\end{center}
\end{figure}
\vspace{-0.15in}

In modeling, we used two fitting parameters, $\Delta_v$ for valley
splitting and $\overline{\tau_q}= (\tau_q^+ + \tau_q^-)/2$ for
quantum lifetime. The former parameter determines entirely the
Fourier spectrum of oscillations and nodes location, whereas the
latter one describes the absolute amplitude of oscillations and
its monotonic field dependence. The difference of partial
lifetimes
for each valley, $(\tau_q^{+}
-\tau_q^{-})/\overline{\tau_q} \ll 1$ was used for a fine
adjustment of  the modulation depth at the node and is found to be
a small parameter, $<3\%$, within the explored range of densities.
Two other parameters of the energy spectrum, $g^*(n)$ and $m^*(n)$
were determined in Ref.~\cite{gm} vs carrier density.
It is important that the desired
valley splitting is determined entirely
by the location of the node of oscillations on the magnetic field
scale (as illustrated in
Fig.~3) and is almost insensitive to $g^*$, $m^*$ and $\tau_q$
values.

We emphasize that the node of beating demonstrated in Figs.~2 and 3 is the first node;
this circumstance enabled
us to determine unambiguously the valley splitting.
The second node might be expected at  magnetic fields which are at least 3 times lower;
however, the  oscillations vanished earlier.

We obtained more than a satisfactory fit using
Eqs.~(3) and (4) for modeling over a whole range of densities where
the beatings are observed; an example  is presented in Fig.~3.
The main result of our fitting,
valley splitting for two samples is presented in Fig.~4.
$\Delta_v$ appears to be sample
dependent, even for the samples with similar peak mobility (same
disorder).

As seen from Fig.~4, over the explored interval of densities,
$n=(7.5-3)\times 10^{11}$\,cm$^{-2}$,
the changes in $\Delta_v$ are about a factor of 1.15 -- 1.3,
much bigger than the error bar, $\sim (2-5)\%$.
The changes though are not reminiscent at all of a
critical behavior, $\Delta_v \propto 1/(n-n_0)$,
which might be expected for the spontaneous valley
polarization at a certain density $n_0 \approx n_c$.
It is noteworthy that over the same interval of densities
both, the electron effective mass $m^*$
and $g^*$-factor were found \cite{gm} to change monotonically
by  a factor of about 1.2.
\vspace{0.4in}
\begin{figure}
\begin{center}
\includegraphics[angle=0,width=3.0in]{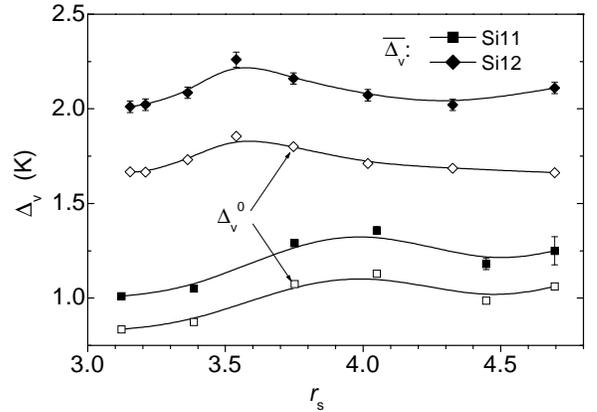}
\begin{minipage}{3.2in}
\vspace{0.1in}
\caption{Valley splitting versus carrier density for two samples.
Filled symbols are for  $\overline{\Delta_v}$,
open symbols
 are  for $\Delta_v^0$.
The uncertainty is represented either by error bar
or the symbols size.}
\label{fig4}
\end{minipage}
\end{center}
\end{figure}
\vspace{-0.2in}

Our measurements were performed in very low though finite fields, $B\approx 0.3$\,T.
It is not clear whether the empiric field dependence
Eq.~(2) remains valid down to $B=0$
or $\Delta_v$ decreases more rapidly  as
levels starts overlapping in low fields.
The accuracy of our data was insufficient to
determine this experimentally though we found the fitting
to be better with field
dependent $\Delta_v(B)$ (as in Eq.~(2)). Given
Eq.~(2) is applicable at $B\rightarrow 0$, it  contributes about
20\% to the obtained $\Delta_v$ values.
We  present therefore in Fig.~4 the results of both fitting, with field independent
$\overline{\Delta_v}$ and with field dependent $\Delta_v=\Delta_v^0 +0.6B$.
The two results have  the meaning of the upper  and
lower estimates, correspondingly, for the true
$\Delta_v(B=0)$ value; their difference is about equal to $0.6 B_{\rm node}$ [K/T],
where $B_{\rm node}$ is the location of the 1st node of oscillations on magnetic field.
The sought-for $\Delta(B=0)$ value is thus within the interval
from $\overline{\Delta_v}$ to $\Delta_v^0$.  Both, $\overline{\Delta_v}$ and
$\Delta_v^0$ exhibit a weak non-monotonic dependence whose origin
is unclear. Our analysis of the ShdH oscillations in the
two-valley system performed up to $r_s=5$ didn't reveal any
deviation from the LK-formulae.
This justifies the assumption
of the conventional Fermi-liquid behavior which we used in the
analysis above.

An additional important result follows from our fitting.
We found the quantum lifetime $\tau_q$ to be almost the same in
the two electron valleys, the maximal difference being only
$(\tau_q^+-\tau_q^-)/\overline{\tau_q} \approx 3\%$   for the density
of $3\times 10^{11}$\,cm$^{-2}$. The proximity of $\tau_q^{\pm}$ values
manifests in Figs.~2 as deep falls in the oscillations amplitude at the beating nodes.
The transport time (momentum relaxation) for Si-inversion layers
is  very close to $\tau_q$ \cite{gm};
it is therefore very likely that the partial transport times and partial
mobilities in both valleys
are almost equivalent.
This is in contrast to the  big, six-fold,  changes
of the resistivity with temperature at zero field,
$\left(\rho(T_{\rm high})-\rho(T_{\rm low})\right)/\overline{\rho(T)} $
which the sample exhibit at the same density \cite{amp2}.
From the proximity of the quantum life times in both valleys,
we conclude therefore that the {\em inter-valley scattering mechanism {\rm \cite{murzin}}
is not responsible
for the strong metallic-like temperature variation of the
resistivity in Si-inversion layers}.

{\em To summarize}, in high mobility (100)Si-inversion layers,
the system which exhibits very strong `metallic-like' features in conduction,
we observed a novel manifestation of valley
splitting: it causes
unexpected beatings in Shubnikov-de Haas oscillations  in
{\em low magnetic fields},\ \ (0.15 --
0.4)\,T, right on the bound of the negative magnetoresistance peak.
From the beatings pattern of oscillations  we
determined the
 valley splitting $\Delta_v$ in the $B=0$ limit.
We found that
$\Delta_v$ varies with density rather weakly and doesn't display a
critical behaviour in the range of densities $(3 - 7.5)\times
10^{11}$cm$^{-2}$ or $r_s =$ 3 -- 5.
We  determined also the individual
quantum lifetimes, $\tau_q^{\pm}$
in both valleys, which appear to
differ by less than $3\%$. This insignificant difference in
$\tau_q$ demonstrates that the semiclassical mechanism of mobility changes
related to the inter-valley scattering
is not the origin of the strong metallic-like temperature dependence of the
resistivity in Si-inversion layers. The findings set novel
constraints on the microscopic  models developed to explain the `metallic'-conduction.

Authors are grateful to A.\ M.\ Finkelstein,
M.\ E.\ Gershenson and H.\ Kojima for discussions. Two of us (V.P. and A.P.)  acknowledge
the hospitality of the Lorenz Center for theoretical physics at Leiden University,
where a part of this  work was done in June, 2000.
The work was supported by NSF DMR-0077825, FWF (13439),
INTAS, German Ministry of Science (DIP),
 RFBR, the Programs \lq Physics of nanostructures',
\lq Statistical physics', \lq Integration' and `The State support of the
leading scientific schools''. A.\ P. acknowledges the support from
Feinberg Graduate School of Weizmann Institute of Science.

\newpage
\section{Appendix: Energy spectrum in Si-inversion layer at low magnetic fields}
The results on the valley splitting presented above shed a light
on the energy spectrum of the two-valley electron system in
(100)Si-inversion layers in {\em low fields}.
We plotted in Fig.~5 the energy for
three lowest Landau levels, $N=0, 1, 2$,  according to
Eqs.~(1) and (2) (from the main section). In high fields, the cyclotron splitting
is the largest and the sequence of energy levels corresponds to
the inset to Fig.~1. Counting from the lowest  Fermi energy,
the 1st gap in the energy spectrum
(filling factor $\nu=1$)
corresponds to valley splitting; $\nu=2$  is for the reduced
Zeeman splitting, $\nu=3$ is again for the valley splitting and
$\nu=4$ is the reduced cyclotron gap. This picture  was verified
in numerous experiments \cite{Aando,AJETP85}.

Due to the nonzero valley splitting at $B=0$,
as magnetic fields decreases,
the energy levels start crossing each other
at  $B \lesssim 1$\,T \cite{quantosc}.
In the region of crossing, this single-electron
picture fails and the repulsion between levels should be
taken into account. There is almost no experimental data available on the
details
of the energy spectrum in Si in low fields. It is well known
only \cite{Adio90,Adio92,AJETPL93} that {\em in low fields/low density regime}
quantum oscillations are missing for $\nu = 3, 4, 5$
and remain pronounced for $\nu=1, 2$  and for $\nu=6$ and
10;  this result was recently confirmed in ref. \cite{misinterp}.

\vspace{0.1in}
\begin{figure}
\begin{center}
\includegraphics[angle=0,width=3.2in]{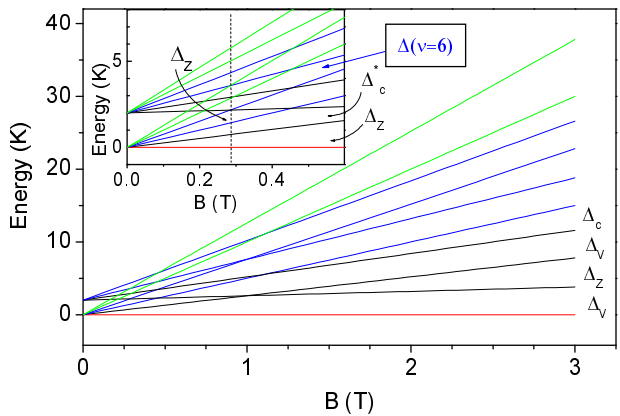}
\begin{minipage}{3.2in}
\vspace{0.1in}
\caption{Schematic magnetic field dependence of the
energy spectrum (in the single-electron approximation) for two
valley electron system in Si and for three lowest Landau
levels. The parameters of the spectrum correspond to Eqs.~(1) and (2);
$\Delta_v =2$\,K, as for the sample Si12.}
\label{fig5}
\end{minipage}
\end{center}
\end{figure}
\vspace{-0.1in}

 In  fields  $B< 0.7$\,T, the sequence of energy
levels changes as shown in the inset to Fig.~5.
For example, at $B=0.6$\,T,   the 1st splitting,  $\nu=1$,
corresponds to the Zeeman gap,
$\nu=2$ is the  cyclotron gap (reduced by the spin flip energy), $\nu=3$ is a
combination of the valley and cyclotron
splitting etc.
In this simplified picture, at some field values ($B=0.45$\,T and 0.25\,T in the inset to
Fig.~5) a number of energy levels coincide.
We don't think though that the
simplified single-electron picture of Fig.~5
may predict exact location of the level crossing regions.

The identification of each energy splitting
in low field/low density regime
is not transparent and requires more thorough theoretical calculations
of the spectrum and much more detailed experiments.
As an example, on the $n-B$ plane in Fig.~6  we reproduce from
Ref.~\cite{quantosc}
a  Landau fan diagram of oscillations in the low density limit;
similar data were reported recently in Ref.~\cite{misinterp}.
In Fig.~6, the $\rho_{xx}$-minima  at $\nu=6$
can be traced without interruption down to the field $B=0.55T$, through the
`metal-insulator' boundary,
however the origin of this splitting may change from `Zeeman gap' at high
fields to any other one, e.g. `valley gap' or `cyclotron gap' at low fields.
By now, there is no firm  physical background to identify $\nu=6$ oscillation
in low fields  with spin splitting.  For example, one can see from the inset to Fig.~5, that
in the single-electron approximation the
$\nu=6$ gap at $B=0.6$\,T  corresponds to
the transition with changing valley (but is reduced by the spin flip energy)
and $\nu=4$ gap is the combined gap
with changing both, the  cyclotron number and the spin projection.

\vspace{0.1in}
\begin{figure}
\begin{center}
\includegraphics[angle=0,width=2.9in]{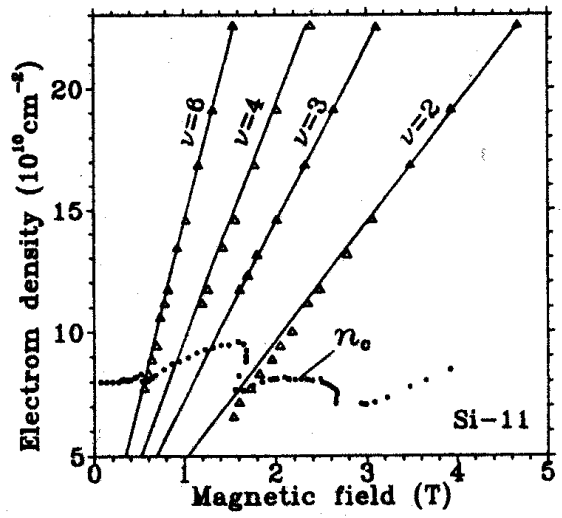}
\begin{minipage}{3.2in}
\vspace{0.1in}
\caption{Landau fan diagram for the four different $\rho_{xx}$-minima,
reproduced from Fig.~2 of Ref.~\protect\cite{quantosc}.
Full dots show MIT-boundary,
defined according to vanishing activation energy.
Other symbols designate location of the $\rho_{xx}$-minima on the $n-B$-plane.}
\label{fig6}
\end{minipage}
\end{center}
\end{figure}

In Ref.~\cite{misinterp}, from the estimated ratio of the two gaps, at $\nu=6$
(which authors of Ref.~\cite{misinterp} assumed to be the spin-gap)
and at $\nu=4$ (was assumed to be the cyclotron gap),
a conclusion was drawn on the unexpectedly strong enhancement of
the $g^*$-factor at low electron densities.
 Taking into account vanishing $\Delta_c$ in the crossing region and
 ill-defined splittings at low field, we find this argument unjustified.
Recent  {\em direct measurements} of the Zeeman energy in low fields vs carrier density
 \cite{Agersh} show that the Zeeman energy increases smoothly as density decreases down to $n_c$,
in agreement with the anticipated
Fermi liquid  renormalization and  with recent results by Okamoto et
al. \cite{Aokamoto}, without any unexpected deviations or divergence
in the range of densities  $n= (1- 90)\times
10^{11}$cm$^{-2}$.

It is noteworthy that for the weak field interference
pattern which we analyzed
above, the interpretation of each oscillation is not
important and does not influence on the conclusion about the
valley origin of the interference and on the results of our analysis.

\end{multicols}

\end{document}